\documentclass[conference]{IEEEtran}
\IEEEoverridecommandlockouts

\usepackage{cite}
\usepackage{amsmath,amssymb,amsfonts}
\usepackage{algorithmic}
\usepackage{graphicx}
\usepackage{textcomp}
\usepackage{xcolor}
\usepackage{csquotes}
\graphicspath{{figures/}} % Set the default folder for images
\usepackage{multirow}
\usepackage{subfig}

\usepackage{tabularx}
\usepackage{caption}
\def\BibTeX{{\rm B\kern-.05em{\sc i\kern-.025em b}\kern-.08em
    T\kern-.1667em\lower.7ex\hbox{E}\kern-.125emX}}
\newcommand\blfootnote[1]{%
	\begingroup
	\renewcommand\thefootnote{}\footnote{#1}%
	\addtocounter{footnote}{-1}%
	\endgroup
}

\makeatletter
\def\footnoterule{\kern-3\p@
	\hrule \@width 3.5in \kern 2.6\p@} % the \hrule is .4pt high
\makeatother
\begin{document}

\title{A Distributed Intelligence Architecture for B5G Network Automation%\\
}
\author{
	\IEEEauthorblockN{Sayantini Majumdar\IEEEauthorrefmark{1}\IEEEauthorrefmark{2}, Riccardo Trivisonno\IEEEauthorrefmark{1}, Georg Carle\IEEEauthorrefmark{2}}
	\IEEEauthorblockA{\IEEEauthorrefmark{1}Munich Research Center, Huawei Technologies
		}
	\IEEEauthorblockA{\IEEEauthorrefmark{2}Technical University of Munich, Germany
		}
	\IEEEauthorblockA{email:[sayantini.majumdar, riccardo.trivisonno]@huawei.com, carle@net.in.tum.de}
}
%\author{Sayantini~Majumdar,
%	Riccardo~Trivisonno,
%	and~Georg~Carle
%\thanks{M. Shell was with the Department
%	of Electrical and Computer Engineering, Georgia Institute of Technology, Atlanta,
%	GA, 30332 USA e-mail: (see http://www.michaelshell.org/contact.html).}% <-this % stops a space
%\thanks{J. Doe and J. Doe are with Anonymous University.}}% <-this % stops a space}% <-this % stops a space

\maketitle
\blfootnote{This work has been submitted to the journal IEEE Networking Letters for possible publication. Copyright may be transferred without notice, after which this version may no longer be accessible.}
\begin{abstract}
The management of networks is automated by closed loops. Concurrent closed loops aiming for individual optimization cause conflicts which, left unresolved, leads to significant degradation in performance indicators, resulting in sub-optimal network performance. Centralized optimization avoids conflicts, but impractical in large-scale networks for time-critical applications. Distributed, pervasive intelligence is therefore envisaged in the evolution to B5G networks. In this letter, we propose a Q-Learning-based distributed architecture (QLC), addressing the conflict issue by encouraging cooperation among intelligent agents. We design a realistic B5G network slice auto-scaling model and validate the performance of QLC via simulations, justifying further research in this direction.
\end{abstract}


\section{Introduction}\label{sec:intro}
Network management automation, often known to diminish the potential for errors by reducing manual intervention, is a significant driver for the development of the next generation of mobile networks \cite{MANO_5G_Americas}. 
%Automation speeds up the time-to-market, reduce operational expenditures (OPEX) and capital expenditures (CAPEX) and improves network performance.  
%The interest in automation is evident from the recent progress made in the telecommunications milieu. 
%E.g., the Zero-touch network \& Service Management (ZSM) provides fully flexible Virtual Network Function (VNF) provisioning services to cope with the immense increase in network traffic and heterogeneous network slices such that the time-to-market and OPEX are significantly reduced\cite{ZSM_001}. 
%In addition, the Service Management and Orchestration (SMO) defined by the O-RAN alliance provides efficient resource management services (scale in and scale out) to the resources of the cloud computing platform, O-Cloud \cite{ORAN_v01}.   
Automation is expected to play a pervasive role in B5G networks, as functionalities of the control plane e.g. the Network Data Analytics Function (NWDAF) and the management plane e.g. Management Data Analytics Service (MDAS) composing the 5G Service-Based Architecture (SBA) become more closely intertwined \cite{akyildiz20206g}. 

In these highly complex networks, automation will be achieved by multiple autonomous, closed loops (CLs) operating concurrently, often on heterogeneous managed objects in different domains -- network functions, network slice instances, access nodes and so on. These autonomous CLs, with predefined individual objectives, often share underlying resources -- thereby affecting the actions of one another. Consequently, the autonomy of these CLs introduces the issue of \textit{conflicts}. A conflict among two or more closed loops may arise when the result of the action of one CL negates or interferes with the result of another. When conflicts are left unresolved, they greatly degrade network performance indicators and stability, thereby negating the gain achieved from automation \cite{hamalainen2012lte}. 
The problem of uncoordinated closed loop actions is even more dire in B5G networks, threatening the smooth evolution of network automation.

Existing research efforts, e.g. in Self-Organizing Networks (SON) in 5G \cite{rojas2019machine}, provides evidence that centralized orchestration avoids the issue of conflicts entirely, as a single entity performs the decision-making. 
However, a centralized approach will not be feasible when there exists an inherent high degree of architectural complexity with which these CLs operate. 
E.g. applications with strict deadlines on optimal management decisions, such as Ultra Reliable Low Latency Communication (URLLC), would be infeasible in a centralized paradigm, as the risk of violating service requirements due to increased signaling overhead would be high \cite{ZSM_005}. 

In this letter, we explore a distributed approach to automating network management decisions, congruent with the envisioned decentralization in B5G networks. Partially inspired by \cite{jiang2018graph}, we propose a solution architecture, Q-Learning for Cooperation ($\mathtt{QLC}$), that consists of a set of autonomous agents, each having a Q-network as its intelligence and operating on its environment. Each agent upon its state space takes actions to reach its individual objective. Its neighbors are other agents with which it shares resources, thereby making resource allocation conflicts imminent. $\mathtt{QLC}$ empowers these autonomously operating agents, by means of essential information exchange of its neighbor agents' variables, to \textit{learn} to take decisions in cooperation with others while attempting to reach the optimal performance. Therefore, the agents are independent learners \cite{claus1998dynamics}, with awareness of neighboring agents' variables to enrich their state space. 
We apply $\mathtt{QLC}$ to a topical B5G case study, auto-scaling, that adjusts shared virtual computing (i.e. CPU) resources to serve incoming network slice load and optimize resource utilization. Results show that $\mathtt{QLC}$ achieves a significant gain over the baseline threshold-based mechanism, similar to the one investigated in \cite{tang2015efficient}. In addition, we show that $\mathtt{QLC}$ performs close to the optimum, achieved by centralized orchestration while minimizing conflicts and that the learning of $\mathtt{QLC}$ agents is robust under dynamic incoming load conditions. Additionally, we observe that $\mathtt{QLC}$ provides an improvement in terms of resource efficiency over the baseline.

Our contributions are as follows: 1) we propose a novel distributed solution architecture, Q-Learning for Cooperation ($\mathtt{QLC}$), to drive the management of B5G networks, while factoring in the issue of conflicts typical of decentralization and 2) we demonstrate the performance gain of $\mathtt{QLC}$ via simulations by applying it to a B5G auto-scaling in network slicing use case.

\section{Related Work}\label{sec:sota}

As of today, there exists little work on distributed architectures factoring in the issue of conflict to advance network automation in B5G. 
\cite{qin2019learning} proposes a QL algorithm \textit{QSON} to solve the conflict arising between Mobility Load Balancing and Energy Saving Management SON functions. Although this paper improves network utility value for different QoS and different time scales, a clear drawback is the \textit{QSON} algorithm components tailored to the specific SON use case. Additionally, \cite{kaur2020energy} formulates an optimization problem to maximize energy efficiency by proposing decentralized, cooperative, multi-agent model-free (QL and SARSA) reinforcement learning schemes. It is, however, not clear how the agents would perform under dynamic system conditions.
Recently, 
%as prototypes evolve into live deployments, operational conflicts are finally being acknowledged. 
ETSI's Zero Touch architecture \cite{ZSM_005} has emphasized the need to avoid \enquote{centralization} of the coordinating entity, e.g. by proposing a static conflict map derived from SON specifications -- for detecting CL conflicts.
%, calculating their optimal behavior and instantiating coordination mechanisms to ensure network stability. 
However, it is unknown how these conflicts would be mitigated after their detection. 

%From the above discussion, we identify significant challenges in coordinating the actions of multiple learning agents -- size of the state-space (the curse of dimensionality), the exploration-exploitation trade-off and existence of multiple equilibrium points, also outlined in \cite{busoniu2008comprehensive}. 
%To this end, recent research focuses on addressing these issues and simplifying the state space to achieve convergence faster. 
\cite{jiang2018graph} proposes a multi-agent cooperative decentralized Q-Learning approach based on graph convolution. Interestingly, it shows that by embedding additional contextual information of neighboring agents in the learning of each agent, cooperation between agents can be achieved. With \cite{jiang2018graph} serving as partial inspiration for our work, we propose $\mathtt{QLC}$ to encourage cooperation in independent agents, thereby enabling advancement towards B5G networks. 

The preceding review substantiates the fact that QL is a useful technology applied to coordinating distributed learning agents. The reason is attributed to the iterative and model-free nature of the QL updates, which means that the agent does not directly learn how to model the environment, rather builds experience by estimating the Q-values using the Bellman Equation \cite{sutton2018reinforcement}. In addition, the agent learns its environment by using sampling policies such as $\epsilon$-greedy approach, wherein it explores by sampling some non-optimal policies of the environment. This strategy, also known as \textit{off-policy} method, enables the agent to not only converge to the optimal action, but also verify that other actions are sub-optimal. 
Based on the terminology of QL in the literature, we categorize QL agents in our solution as independent learners (i.e. independent action space) \cite{claus1998dynamics} with the novelty of neighbor information exchange embedded in the state formulation to induce cooperation. 
\section{Q-Learning for Cooperation ($\mathtt{QLC}$)}\label{sec:concept}
This section proposes a decentralized approach to network automation, to leverage the gain of decentralization, while addressing the critical issue of conflicts which may occur due to the concurrent operation of CLs. 
\subsection{Our proposed $\mathtt{QLC}$ architecture}\label{sec:concept_general}
%The proposed approach exploits Q-Learning (QL) to encourage cooperation among autonomous Intelligent Agents (IAs) aiming to reach their local objectives while still factoring in the issue of conflicts. 
%\begin{figure}
%	\centerline{\includegraphics[width=0.7\linewidth]{fig1.eps}}
%	%\centerline{\includegraphics[page=2,width=\linewidth]{figures_1.pdf}}
%	\caption{$\mathtt{QLC}$ reference architecture}
%	\vspace*{-5mm}
%	\label{fig:CL_neighbors}
%\end{figure}
We consider an automated system 
%in Fig.~\ref{fig:CL_neighbors} 
that constitutes $N$ independent closed loops (CLs), each managed by an Intelligent Agent (IA) $A_{i}(1< i \le N)$, empowered with QL capabilities. Each IA $A_i$ implements the CL upon observation of a set of $n_i$ local variables $x_{i,k}(1\le k \le n_i)$ and taking an action $a_{i,l} \in \mathcal{A}_i (1\le l \le \left \vert \mathcal{A}_i \right \vert)$ in an environment, constituting a CL iteration. This allows each IA to pursue optimization over local variables. 
The number of IAs $D_i \le N-1$ whose actions may impact local variables of $A_i$ are defined as Neighbor Intelligent Agents (NIAs) of $A_i$. We assume that neighboring agents are able to share their knowledge among themselves. Conflicts among different IAs may occur, whenever the actions $a_{i,l}$ of $A_i$ may impact the local variables $x_{j,k}$ of a different IA $A_j$.

At each control iteration, each IA $A_i$ in the $\mathtt{QLC}$ framework determines its state $s_{i,p}$ where $s_{i,p} \in \mathcal{S} (1\le p \le \left \vert \mathcal{S} \right \vert)$. $\mathtt{QLC}$ encourages each IA $A_i$ to select its actions by embedding in $s_{i,p}$ the impact of its local variables $x_{i,k}$ as well as those of its $D_i$ neighbors $x_{D_i,k}$. The core idea of this approach allows $A_i$ to learn independently i.e. in a distributed manner, but with cooperation embedded in $s_{i,p}$ of $A_i$, thereby avoiding an increase of $\mathcal{A}_i$. 
%In Fig.~\ref{fig:CL_neighbors}, $A_2$ is the neighbor to $A_1$ and $A_3$, implying the meaning of the arrows. 
%The criteria for configuring neighbor relationships is subject to future work.

Each IA $A_i$ stores a $ \left \vert \mathcal{S} \right \vert \times  \left \vert \mathcal{A}_i \right \vert$ Q-table $Q_i$, representing a function $Q_i:s_{i,p}\times a_{i,l}  \rightarrow \mathbb{R}$. Each cell of $Q_i$, also called the action value function $Q_i(s_{i,p}, a_{i,l})$, represents the expected long-term rewards corresponding to each state-action pair. 
%$Q_i$ allows the selection of the best action corresponding to current state $s_{i,p}$ (i.e. the action maximizing the reward). 
After an action has been taken, impacting local and neighbor variables, the state may change. The IA assesses the reward $r$ of the action taken and updates $Q_i$ according to the Bellman Eqn. \cite{sutton2018reinforcement} in (\ref{eq:Bellman}) below. 
%\vspace{-4mm}
\begin{multline}\label{eq:Bellman}
Q_i(s_i,a_{i,l}) \leftarrow Q_i(s_{i,p},a_{i,l}) + \\
\alpha\left(r(s_{i},a_{i,l})+ \gamma \max_{a_{i,l}}Q_i(s^\prime_i, a_{i,l})- Q_i(s_i,a_{i,l})\right),
\end{multline}
where $\alpha$ is the learning rate and $\gamma$ is the discount factor. Here, $s_i$ and $a_{i,l}$ are the current state and action respectively, while $s^\prime_i$ is the new state which action $a_{i,l}$ brings the agent to. 

The learning principle is grounded in the two phases of Q-learning: exploration and exploitation. 
%$Q_i$ is initialized with dummy values which do not privilege any action over another. 
Using the $\varepsilon$-greedy approach \cite{sutton2018reinforcement}, exploration allows an IA with a probability $\varepsilon$ to randomly select actions, sampling both optimal and sub-optimal actions, evaluating and updating the quality of the action according to Eqn.~(\ref{eq:Bellman}). An IA is considered to have explored long enough once its Q-values $Q_i(s_{i,p}, a_{i,l})$ do not exhibit substantial changes any more and it is ready to exploit its learned knowledge, i.e. when in state $s_{i,p}$, action $a_{i,l}$ is selected as $\max_{a_{i,l}}Q_i(s^\prime_i, a_{i,l})$.
\subsection{$\mathtt{QLC}$-based auto-scaling system model} 
\begin{figure}
	\centerline{\includegraphics[width=1.0\linewidth]{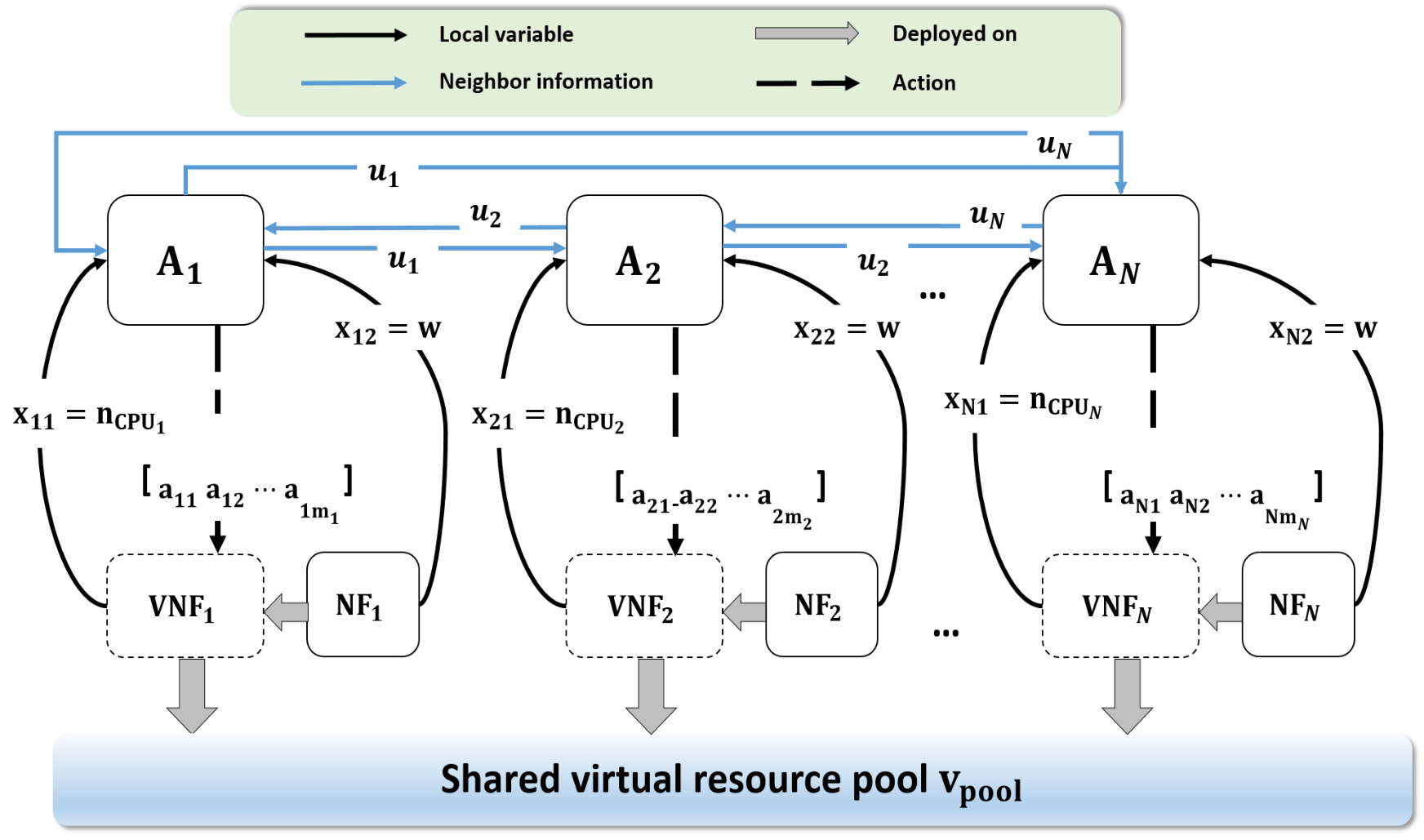}}
	%\centerline{\includegraphics[page=2,width=\linewidth]{figures_1.pdf}}
	\caption{Network function auto-scaling system model}
	\vspace*{-5mm}
	\label{fig:exp_scenario}
\end{figure}
To validate the proposed framework $\mathtt{QLC}$, in this letter we investigate its application to auto-scaling, a relevant B5G resource orchestration use case.

We consider a virtualized environment consisting of a Network Slice (NS) composed of Network Functions (NFs), where each NF is implemented as software on a Virtual Network Function (VNF). These VNFs share a virtual computing (i.e. CPU) resource pool, hosted on physical infrastructure via a virtualization layer \cite{ETSI_NFV-MAN_001}.  
At time instant $t$, a population of User Equipment (UEs) may issue service requests to the NS. The auto-scaling mechanism monitors the number of UEs admitted by the NS, $w(t)$, that represents the load generated to the NF and the number of CPUs allocated to its VNF $n_{CPU_i}(t)$. The actual load generated by the $k^{th}$ UE is $\mu_{UE_k}(t)$. 
The VNF CPU utilization $u_i(t)$ is computed as directly proportional to $\sum_{k=1}^{w(t)}\mu_{UE_k}(t)$ and inversely related to $n_{CPU_i}(t)$. Given the monitored variables, 
auto-scaling regulates the number of virtual CPUs allocated to each VNF according to the incoming NS load, aimed at bringing $u_i(t)$ to a target VNF CPU utilization $u_T$. $u_T$ is defined according to resource efficiency and slice reliability criteria, to avoid under and over provisioning of resources without compromising the ability of the NS to serve incoming load. To achieve this objective, auto-scaling scales down CPU resources when NS load is low and scales up when it is high. During high incoming NS load, all NIAs try to scale up CPUs from the same resource pool. When the resource pool is unable to satisfy the combined demand of NIAs, only \textit{one} of the NIAs is privileged while the rest are given no extra CPUs. In our design, we define this event as a conflict.
Evidently, conflicts may result in an unbalanced resource sharing among NF-VNF pairs, affecting the maximum load the NS may serve and causing inefficient resource provisioning. Considering this problem, below we describe our $\mathtt{QLC}$ solution design.

\textbf{Monitored variables. }In addition to monitoring its own variables $w(t)$ and $n_{CPU_i}(t)$, each IA $A_i$ in the $\mathtt{QLC}$ framework collects VNF utilization of each of its $D_i$ neighbors, illustrated in Fig.~\ref{fig:exp_scenario}.

\textbf{State space. }Embedding knowledge of neighbors' variables in the state formulation to encourage cooperation forms the core novelty of our solution. In this regard, the proximity of $u_i$ to $u_T$ and the VNF utilization of $D_i$ NIAs must be assessed in order to select the proper auto-scaling action. Moreover, as the occurrence of conflicts may lead to an uneven resource sharing among VNF-NF pairs, a proper auto-scaling action selection must consider how balanced the load is among the NIAs. To this end, a two dimensional state formalized as a complex variable 
\begin{equation}\label{eq:state_complex}
	s_{i,p} = s_{i,p}^I + is_{i,p}^b
\end{equation}
encodes the two aspects of the state design. Here, 
$s_{i,p}^I \in \{s_{-B}^l, s_{-B+1}^l, …, s_{-1}^l s_0^l, s_1^l, … s_{B-1}^l, s_B^l \}$ 
is a discrete variable representing the degree of loading of $A_i$ with regard to $D_i$ NIAs, which can assume $2B+1$ values. $s_0^l$ is the state where $\left(\frac{u_i+\sum_{k=1}^{D_i}u_k}{D_i+1}\right)$ is minimized.
Moreover, $s_{i,p}^b$ is a discrete variable measuring the balancing among $A_i$ and $D_i$ NIAs, determined by the sign of $\Delta u = \left(u_i - \frac{1}{D_i+1}\cdot\sum_{k=1}^{D_i+1}u_k \right)$ according to the criteria 
\begin{equation}
s_{i,p}^b = 
\begin{cases}
-1; & \Delta u < 0,\\
0; & \Delta u = 0,\\
+1, & \Delta u > 0. 
\end{cases}
\end{equation}
Extended formulas for (\ref{eq:state_complex}) are omitted for brevity.
%A critical issue typically encountered in multi-agent environments is the explosion of the state space with the number of agents. With our composite state encoding strategy, the dimensions of the $Q_i$ maintained by each IA would be $\left((2B+1)\times 3\right) \times \left \vert \mathcal{A}_i \right \vert$. Such a representation evades an exponential increase in the number of states with the number of agents.

\textbf{Action space. } The action space of $A_i$ is a discrete, finite set denoted by $\mathcal{A}_i \subseteq \mathbb{Z}$ where $\mathbb{Z}$ is the set of integers. 
%The attempt of increase may lead to conflict and hence fail in case of insufficient resources.

\textbf{Reward model. }Two aspects need to be accounted for in the design of the reward function $r_i$. First, each IA adjusts the number of CPUs aiming to reach the target utilization $u_T$. Hence, the closer an action brings the utilization to $u_T$ the higher the action shall be rewarded. Second, actions incurring in conflicts shall be penalized. As conflicts ultimately lead to the number of CPUs to remain unchanged after the attempted action (except for the privileged IA), all NIAs associated with the conflict will be penalized, according to the formula
\begin{equation}\label{eq:reward} r_i = 
\begin{cases}
	c\cdot K\cdot(\left \vert u_i - u_T \right \vert - \left \vert u^{\prime}_i - u_T \right \vert); & \text{if $a_{i,l} \ne 0$}, \\
	\frac{(u_{T})^2}{(u_i - u_{T})^2 + \delta^2}; & \text{otherwise},
\end{cases}
\end{equation}
where $u^{\prime}_i$ is the updated utilization after executing action $a_{i,l}$, $c$ is a flag to determine conflict, $K$ is a constant to shape the reward and $\delta$ is a constant to avoid singularities.	
%In summary, an agent in our proposed cooperative framework aims to reach as close as possible to its own optimal performance while resolving conflicts and finding the most efficient way of sharing resources with its neighbors. 

\section{Experimental evaluation}\label{sec:evaluation}
The experimental evaluation aims at exploring potential gains of the proposed Q-Learning algorithm for Cooperation, or $\mathtt{QLC}$, and compares performance with an existing auto-scaling mechanism investigated in \cite{tang2015efficient} as well as with a centralized orchestration achieving the theoretical optimal solution.
\begin{table}
	\caption{Simulation configuration parameters}
	\footnotesize
	\begin{center}
		\begin{tabularx}{\linewidth}{|p{0.7cm}|p{3.3cm}|l|p{1.4cm}|p{0.35cm}|}
			\hline
			\textbf{Type} & \textbf{Parameter} & \textbf{Symbol} & \textbf{Value} & \textbf{Unit} \\
			\hline
			\multirow{6}{*}{System} & Admission control threshold & $AC_{thr}$ & 0.9 & -\\
			& Scale-up threshold & $SC_{high}$ & 0.95 & -\\
			& Scale-down threshold & $SC_{low}$ & 0.15 & -\\
			& CPU utilization target & $u_T$ & 0.5 & -\\
			& Initial no. of CPU per VNF & $n_{CPU_i}$ & 1 & -\\
			& No. of available CPU & $v_{pool}$ & 20 & -\\
			& Episode duration & $T$ & $10^5$ & s\\ 
			& No. of episodes & $E$ & 20 &-\\
			\hline
			\multirow{4}{*}{Load} & Population of users & $U$ & $10^5$ & - \\
			& Service request/user & $\lambda_{UE}$ & $5\times 10^{-7}$ - $2\times 10^{-5}$ & s$^{-1}$ \\
			& Service duration (mean, sd)& $\bar{\theta},\sigma_{\theta}$ & 60, 5 & s \\
			& Actual load/user (mean, sd)& $\bar{\mu}, \sigma_{\mu}$ & 1, 0.02 & - \\
			\hline
			\multirow{3}{*}{Agent} & Learning rate & $\alpha$ & 0.5 & -\\
			& Discount factor & $\gamma$ & 0.9 & - \\
			& $\epsilon$ initial, final & $\epsilon_{i}, \epsilon_{f}$ & 0.9, 0.0001 & - \\
			\hline
		\end{tabularx}
		\vspace*{-5mm}
		\label{tab:sim_params}
	\end{center}
\end{table}
	\subsection{Simulation setup}
	The auto-scaling algorithms are applied to a system consisting of an NS composed by two NFs, $NF_1$ and $NF_2$ placed on a VNF each, outlined in Table~\ref{tab:sim_params}. 
	%Each NF is managed independently by an agent, either implementing $\mathtt{QLC}$ or sending/receiving management commands to or from the centralized orchestrator. 
	The NS implements a distributed threshold-based admission control, allowing load to be admitted if utilization of the $i^{th}$ VNF does not exceed a local threshold $AC_{thr_i}$. Additionally, the NS is initially configured with a number of available virtual CPUs $v_{pool}$ while $n_{CPU_1}$ and $n_{CPU_2}$ number of initial CPUs are allocated to $NF_1$ and $NF_2$ respectively. In our evaluation, we configure the auto-scaling actions that $A_1$ and $A_2$ may attempt at every CL iteration to reach $u_T$ to be: an increase or decrease of one or two CPUs, or maintaining the number of CPUs unchanged. Hence, the action set of each agent $A_i$ in our evaluation is $\mathcal{A}_i=\{-2,-1,0,+1,+2\}$. 
	%Performance of the algorithms is dependent on initial conditions and would vary based on these assumptions.
	
	\textbf{User model setup.} Arrivals of UEs to simulate loading the NFs are modeled in two sets of scenarios. In Scenario 1, the NS is loaded by service requests coming from a population of UEs $U$, with Poisson arrival rate per UE $\lambda_{UE}$, a service duration $\theta$ and a generated load $\mu$ modeled as Gaussian variables. The incoming load generated to the NS is $\Lambda_{in}$ the aggregate arrival rate of $U$ UEs each with $\lambda_{UE}$ arrival rate. 
	Next, Scenario 2 replicates a dynamic environment according to the principles of a realistic diurnal scenario from \cite{wang2015understanding}, with arrival rate per UE $\lambda_{UE}(t)$ varying dynamically in time. We also define an episode as a complete simulation duration from $t=0s$ to $t=Ts$. Owing to the stochastic nature of our simulations and to enable the IAs to learn the dynamic environment, we implement Q-table learning over multiple episodes.
	%, by initializing the input $Q_i$ of the next episode with the final $Q_i$ outputs from the previous. 
	We configure an episode duration $T=10^6s$ simulating $\sim$27.7 hours of service requests, constituting smooth increase and decrease over two $\sim$13.9 hour periods. The two peaks of incoming service requests reflect the periods of peak activity over $T$ across little more than a 24-hour period.  
	
	We evaluate the system performance by examining a number of metrics and events. First, the ability of the network slice to serve the incoming load $\Lambda_{in}$ is measured by $\Lambda_{out}$. Further, we consider the ability of the management system to ensure a VNF utilization close to the target $u_T$, which is regarded as a resource efficiency metric. Therefore, we define the Resource Efficiency Indicator ($REI$) \vspace{-3mm}
	\begin{equation}\label{eq:REI}
	REI = \frac{1}{N}\cdot\sum_{i}^N\frac{u_i}{u_T}.
	\end{equation}
	Finally, the occurrence of conflicts is also treated as an empirical performance indicator.  
%		\begin{figure*}
%		\centering
%		\vspace{-20pt}
%		\subfloat[Episode 1]{\def\svgwidth{0.5\columnwidth}
%			\input{figures/hmap_Agent0_ep01.pdf_tex}}
%		%\vspace{-0.5em}
%		%\hspace{0.5em}
%		\subfloat[Episode 10]{\def\svgwidth{0.5\columnwidth}
%			\input{figures/hmap_Agent0_ep10.pdf_tex}}
%		%\vspace{-0.5em}
%		%\hspace{1.5em}
%		\subfloat[Episode 20]{\def\svgwidth{0.5\columnwidth}
%			\input{figures/hmap_Agent0_ep20.pdf_tex}}
%		\caption{$Q_1$ heat maps across episodes}
%		\vspace{-20pt}
%		\label{fig:heat_map}
%	\end{figure*}

	We benchmark system performance by defining the no auto-scaling mechanism $\mathtt{NO\_AUT}$, where $n_{CPU_i}$ remain unchanged throughout the simulation, serving as the lower bound with the given infrastructure settings.
	The threshold-based auto-scaling mechanism $\mathtt{THR}$ employs a greedy (non-cooperative), distributed approach to pursue $u_T$, by triggering scale up actions when $u_i$ exceeds a congestion threshold $SC_{high}$ and releases CPUs when $u_i$ falls below a resource under-utilization threshold $SC_{low}$.
	In addition, we formalized an Mixed Integer Optimization ($\mathtt{MIO}$) formulation, implemented in a VNF orchestrator, aiming at the optimal CPU allocation to VNFs, with an objective function maximizing the served load and minimizing the differences between $u_i$ and $u_T$. Evidently, at each CL iteration, the $\mathtt{MIO}$ formulation entails high signaling to collect $u_i$ from all VNFs to command CPU adjustment and high computation power. 
	%Therefore, the results from $\mathtt{MIO}$ set the theoretical optimum we seek. 
	The $\mathtt{MIO}$ problem formulation is omitted for brevity. 

	\subsection{Simulation results}
	We evaluate the performance of $\mathtt{QLC}$ for $N=2$ IAs, using configuration parameters in Table~\ref{tab:sim_params}. 
	%The performance metrics that are evaluated include the comparison of $\Lambda_{out}$ and $REI$ for all aforementioned algorithms $\mathtt{NO\_AUT}$, $\mathtt{THR}$, $\mathtt{QLC}$ and $\mathtt{MIO}$. 
\begin{figure}
	\centerline{\includegraphics[width=0.8\linewidth]{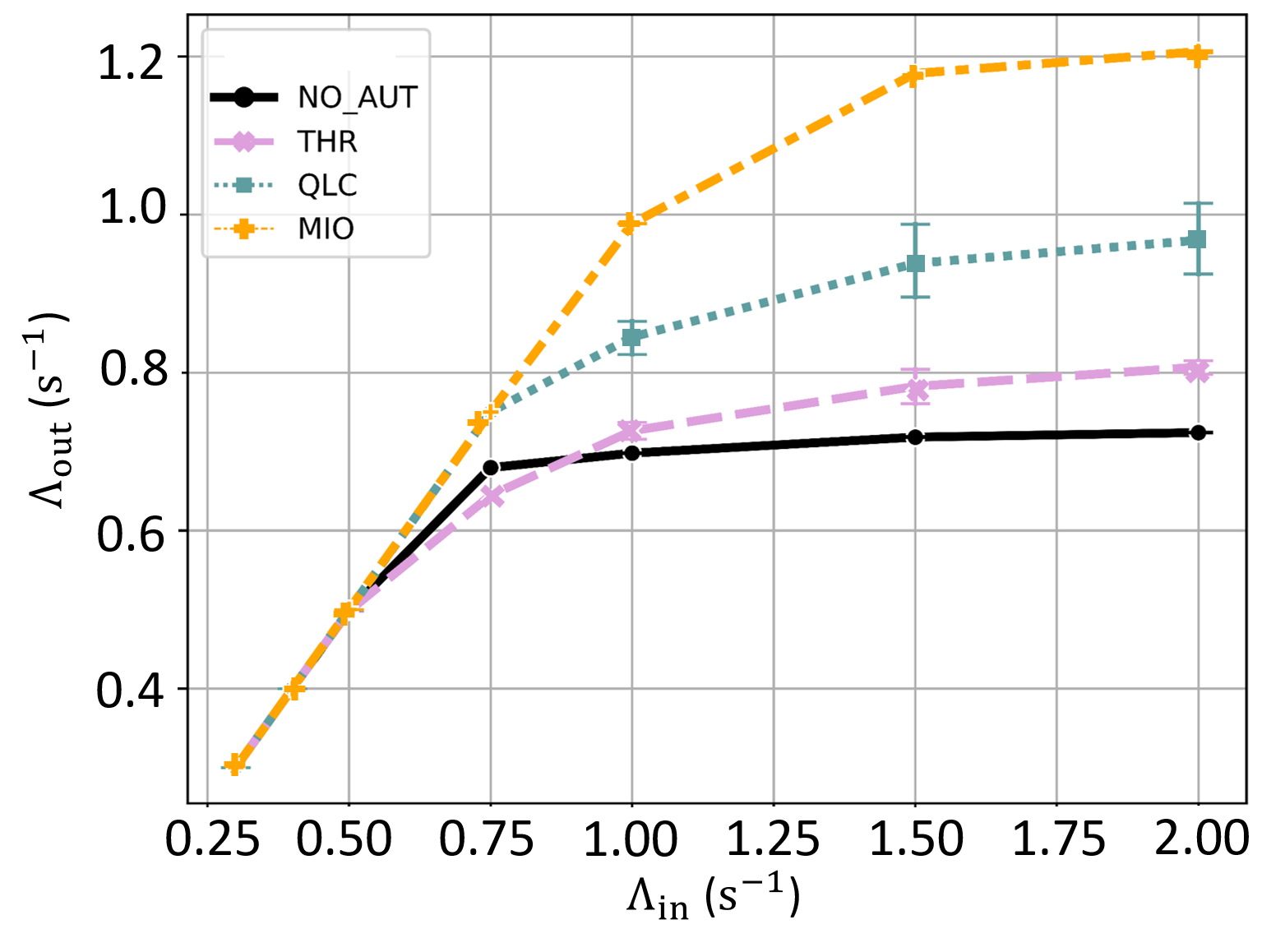}}
	%\centerline{\includegraphics[page=4,width=\linewidth]{figures_1.pdf}}
	\caption{Served load $\Lambda_{out}$ (s$^{-1}$)}\label{fig:served_load_all}
	%\vspace*{-5mm}
\end{figure}

	\textbf{Served load.} A 95\% confidence interval plot of $\Lambda_{out}$ vs. $\Lambda_{in}$ for Scenario 1, shown in Fig.~\ref{fig:served_load_all}, indicates that $\mathtt{MIO}$ provides the optimal CPU allocation a centralized management system may achieve. 
	All algorithms show identical performance at low load, as the initial system configuration resources are sufficient to serve all the load. $\mathtt{QLC}$ determines a clear improvement compared to $\mathtt{THR}$, as saturation effect appears at $\Lambda_{in}=1.0$s$^{-1}$ and $0.75$s$^{-1}$ respectively. $\mathtt{QLC}$ improves the maximum served load, approximately $1.0$s$^{-1}$ vs $0.8$s$^{-1}$ at $\Lambda_{in}=2.0$s$^{-1}$, with a gain of $\sim$25\%. The wide confidence interval at high loads reflects the multi-equilibrium problem that adversely affects the performance of Q-Learning.
	
		\begin{figure*}
		\centering
		\centerline{\includegraphics[width=1.9\columnwidth]{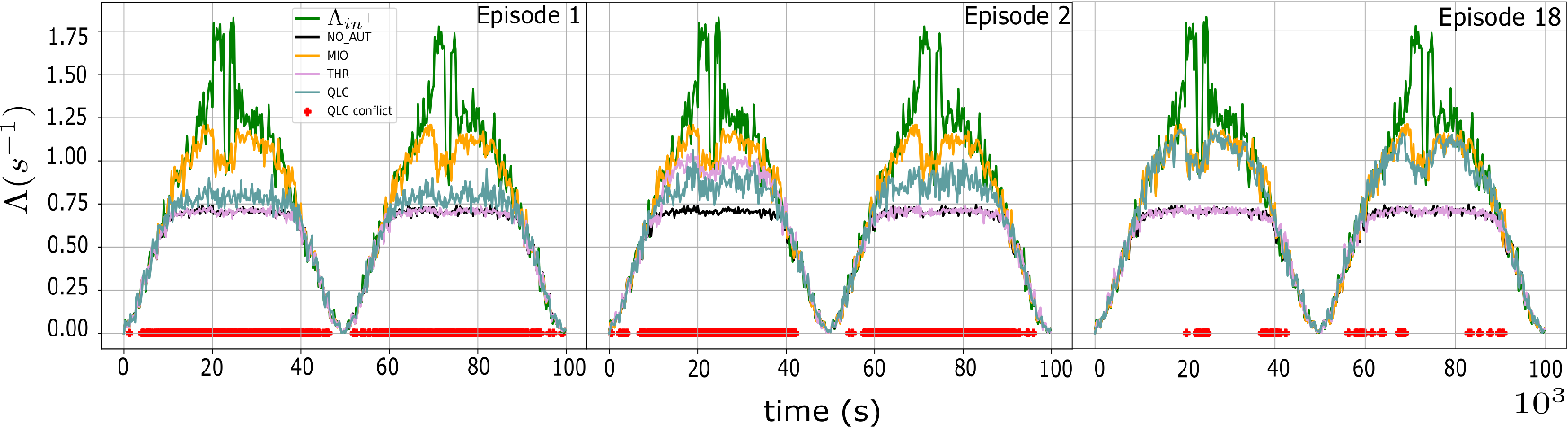}}
		\caption{System load $\Lambda_{in}(s^{-1})$ and corresponding $\Lambda_{out}$ ($s^{-1}$)}
		\label{fig:served_load_dynamic}
		%\vspace*{-5mm}
	\end{figure*}
	Let us consider Scenario 2. Fig.~\ref{fig:served_load_dynamic} depicts the time evolution of system load $\Lambda_{in}(t)$ and $\Lambda_{out}(t)$ for the corresponding algorithms. The timestamps of conflict event occurrences due to $\mathtt{QLC}$ are also highlighted. It is observed that $\mathtt{QLC}$ shows little improvement of $\Lambda_{out}(t)$ in episode 1, as the IAs have just begun sampling non-greedy actions to improve their current estimates of $Q_i(s_{i,p}, a_{i,l})$. Exploration, therefore, drives IAs to record a large number of conflicts in episode 1. $\mathtt{THR}$ shows no apparent gain at certain episodes because $SC_{high}$ is not reached due to the randomness of $\mu_{UE_k}(t)$. In episode 2, $\mathtt{QLC}$ shows a steady increase in $\Lambda_{out}(t)$ compared to episode 1. On the other hand, $\mathtt{THR}$ performs well and even better in the first cycle than $\mathtt{QLC}$ as $SC_{high}$ is reached and $\mathtt{THR}$ scales $n_{CPU_i}$ up to serve more users. However, this apparent improvement is not reliable as $\mathtt{THR}$ performs poorly again in the next cycle. On the other hand, $\mathtt{QLC}$ performs equally well in both cycles. This observation solidifies the importance of learning for the ability of the system to serve more load, the robustness of $\mathtt{QLC}$ and validates that $\mathtt{QLC}$ indeed learns across episodes. After a subsequently high number of episodes, e.g. in episode 18 first we begin to observe $\mathtt{QLC}$ performing quite close to the optimal $\mathtt{MIO}$ while reducing the number of conflicts. These observations show that for $\mathtt{QLC}$ to exhibit near optimal performance while addressing conflicts, a strategy that first allows the IAs to learn for a few days before being deployed could be followed. At certain timestamps, $\mathtt{MIO}$ exhibits greater $\Lambda_{out}(t)$ than $\Lambda_{in}(t)$ due to the granularity of the measurements. 

	\begin{figure}
		\centerline{\includegraphics[width=0.8\linewidth]{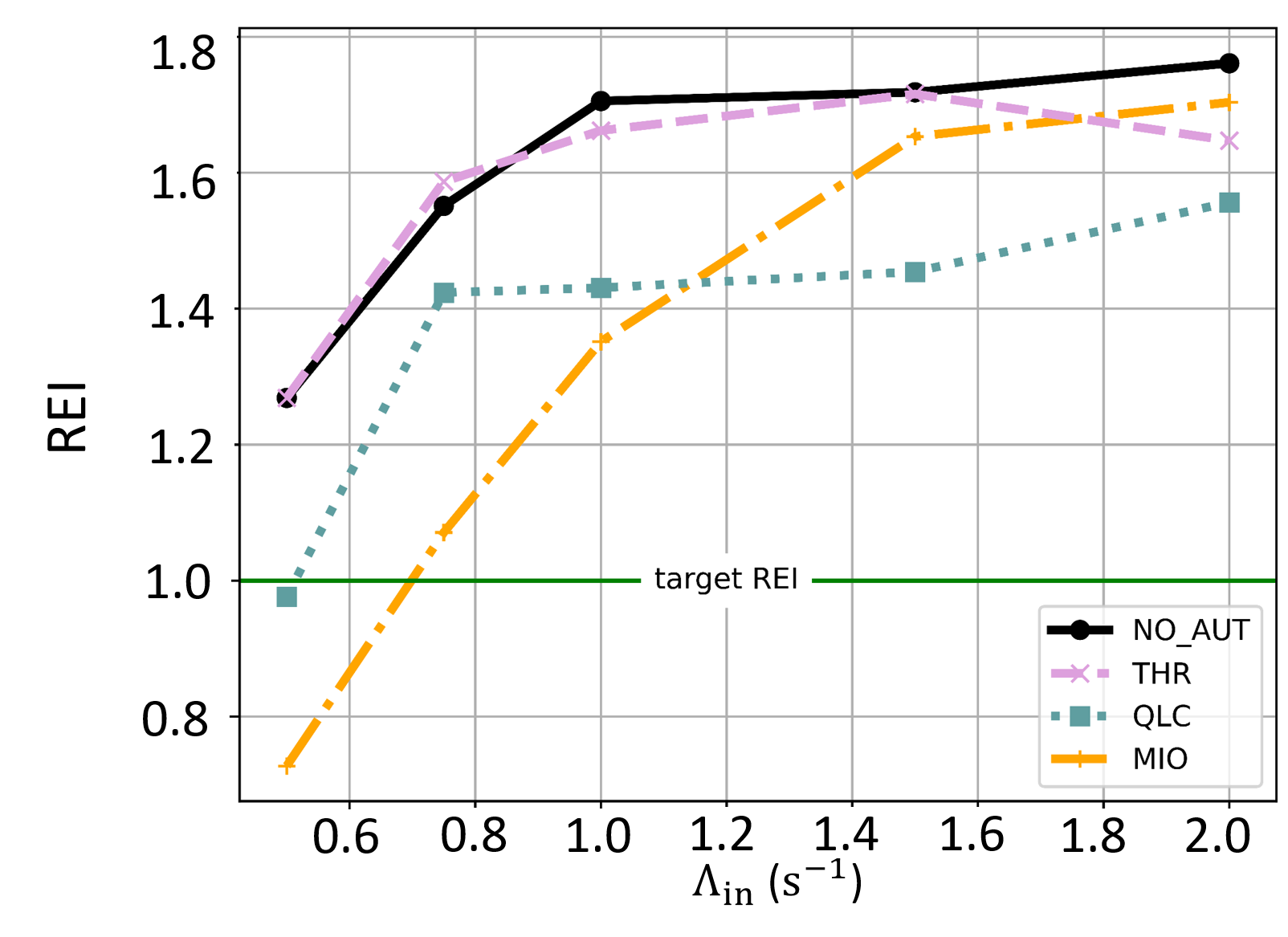}}
		%\centerline{\includegraphics[page=5,width=\linewidth]{figures_1.pdf}}
		\caption{Resource efficiency indicator $REI$}\label{fig:ee}
		%\vspace*{-5mm}
	\end{figure}

	\textbf{Resource efficiency.} Fig.~\ref{fig:ee} illustrates $REI$ vs. $\Lambda_{in}$ for Scenario 1. The gain of $\mathtt{QLC}$ with respect to $\mathtt{THR}$ is of immediate reading. Here, $\mathtt{MIO}$ still represents the bound of optimal performance. The apparent better performance of $\mathtt{QLC}$ at high load is attributed to the ability of $\mathtt{MIO}$ to serve higher load, as observed in Fig.~\ref{fig:ee}.

\section{Conclusions \& Future Work}\label{sec:conclusion}
In this paper, $\mathtt{QLC}$, a decentralized approach to B5G network automation has been proposed, aiming at local optimizations while simultaneously resolving potential conflicts which may arise among concurrent CLs.
%The proposed framework leverages reinforcement learning, by the introduction of an intelligent agent, featuring Q-learning capabilities at the heart of each CL.
The Q-Learning framework $\mathtt{QLC}$ has been applied to the practical problem of NF auto-scaling in a network slice. A detailed design of the solution was proposed. 
Performance is assessed in terms of the maximum load the network slice can serve and resource efficiency, measured by the capability of the network slice to keep CPU utilization close to a target. 
Performance has been compared to an optimal centralized orchestration solution and to an existing auto-scaling mechanism currently implemented in real systems. Simulation results highlight the potential of $\mathtt{QLC}$ which, in the scenarios examined, decreases the occurrence of conflicts after a training period. $\mathtt{QLC}$ achieves to up to $\sim 25\%$ gain compared to the existing decentralized mechanism and would also not incur the drawbacks of the optimal centralized orchestration. Moreover, an analysis of the performance of $\mathtt{QLC}$ agents for dynamic incoming load shows that $\mathtt{QLC}$ is robust even in realistic scenarios.
This seminal work will require massive future analysis towards the definition of pervasive intelligence in B5G networks, e.g. investigating performance and convergence for scenarios with multiple agents, criteria for selecting neighbor groups and even comparison of different reward function formulations.\vspace*{-5mm}

\vspace{12pt}

\bibliographystyle{ieeetr}
\bibliography{bibliography}
\end{document}